\documentclass[review]{elsarticle}
\usepackage{lineno,hyperref}
\modulolinenumbers[5]

\journal{Journal of Nuclear Physics A}

\begin{document}

\begin{frontmatter}

\title{{Measurement of alpha-induced reaction cross-sections for $^{nat}$Zn with detailed covariance analysis}}

\author[1]{Mahesh Choudhary\corref{mycorrespondingauthor}}
\ead{maheshchoudhary921@gmail.com}
\author[1]{Aman Sharma}
\author[1]{Namrata Singh}
\author[1]{A. Gandhi}
\author[2]{S. Dasgupta}
\author[2]{J. Datta}
\author[3]{K. Katovsky}
\author[1]{A.~Kumar\corref{mycorrespondingauthor}}
\ead{ajaytyagi@bhu.ac.in}
\address[1]{Department of Physics, Banaras Hindu University, Varanasi-221005, India}
\address[2]{Analytical Chemistry Division, Bhabha Atomic Research Centre, Variable Energy Cyclotron Centre, Kolkata-700064, India}
\address[3]{Department of Electrical Power Engineering, Brno University of Technology, Brno-61600, Czech Republic}
\cortext[mycorrespondingauthor]{Corresponding author}

\begin{abstract}
The production cross-section of $^{68}$Ge, $^{69}$Ge, $^{65}$Zn and $^{67}$Ga  radioisotopes from alpha-induced nuclear reaction with $^{nat}$Zn have been measured using the stacked foil  activation technique followed by the off-line $\gamma$-ray spectroscopy in the incident alpha energy range 14-37 MeV. The obtained nuclear reaction cross-sections are compared with previous experimental data available in the EXFOR data library, evaluated nuclear data from TENDL-2019 and theoretical results, calculated using TALYS nuclear reaction code. We have also performed the detailed uncertainty analysis for these nuclear reactions and their respective correlation metrics are presented. Since $\alpha$-induced reactions are important in nuclear medicine and developing  the nuclear reaction codes so needful corrections related to the coincidence summing factor and the geometric factor have been considered during the data analysis in the present study.

\end{abstract}
\end{frontmatter}
\section{Introduction}
The data on alpha-induced  nuclear reaction cross-sections is important for a variety of technological applications including nuclear reaction investigations and the production of medical radionuclides \cite{03,04}. Radioisotopes are being used as therapeutics  for a long time and $^{68}$Ga is one of such radioisotopes used for PET (Positron Emission Tomography). The production of $^{68}$Ga has became more available with the increasing number of medical cyclotrons recently. However, $^{68}$Ga has a short half-life of 67.71 minutes and emits positrons with the positron branching ratio 89 $\%$ accompanied by 1077.34 keV $\gamma$-ray. Transport of radioisotopes, like $^{68}$Ga, becomes difficult due to their short half-lives.  So using $^{68}$Ge as a parent to $^{68}$Ga is a feasible solution, as it decays to $^{68}$Ga with 100 $\%$ electron capture along with having a relatively longer half-life of 270.95 days. Hence $^{68}$Ga/ $^{68}$Ge generator is an ideal candidate to be used in distant places from the manufacturing site. There are several reactions to produce $^{68}$Ge like $^{nat}$Zn($\alpha$,x), $^{66}$Zn($\alpha$,2n) etc.\\
Another prominent medical radioisotope is the $^{67}$Ga which is commonly used in nuclear medicine for various types of human tumors and inflammatory lesions \cite{another}. The $^{67}$Ga radioisotope has a half-life of 3.26 days and emits $\gamma$-ray of 300.22 keV. In this study, $\alpha$ particles bombarded on a natural zinc target with an energy of 37 MeV to produce the above-mentioned medical radioisotopes. Since $\alpha$-induced reactions are significant in nuclear medicine and developing nuclear reaction codes, precisely estimating the degree of uncertainty propagation from the measured nuclear reaction cross-section data of these nuclear reactions is an important component. Although the EXFOR library \cite{13,14} contains experimental data for these nuclear reactions, none of the data has a complete covariance analysis. Covariance analysis is a method for estimating the uncertainty in a measured quantity by taking cross-correlations into different attributes \cite{Mahesh,10}. In the present work, we have documented detailed covariance analysis of nuclear reactions $^{nat}$Zn($\alpha$,x)$^{68}$Ge, $^{nat}$Zn($\alpha$,x)$^{69}$Ge, $^{nat}$Zn($\alpha$,x)$^{65}$Zn and  $^{nat}$Zn($\alpha$,x)$^{67}$Ga, by taking the micro correlations between different attributes like decay constant,  incident flux, the efficiency of HPGe detector, $\gamma$-ray counts, $\gamma$-ray intensity and particle number density. We have used TALYS nuclear code \cite{16} for the theoretical calculation of the nuclear reaction cross-section. In the present work, due to the importance of these models, the impacts of six-level density models on the cross-section measurements for the production of the radionuclides $^{68}$Ge, $^{69}$Ge, $^{65}$Zn, and $^{67}$Ga through $^{nat}$Co(a,x) reactions were examined. The presented excitation functions of these nuclear reactions are compared with the existing experimental data available in the EXFOR library, evaluated nuclear data from TENDL-2019, as well as the theoretical calculation.\\
The following six components make up the present study, section 2 contains information on the experimental technique and setup, section 3 covers information on the detector's efficiency calibration, section 4 provides information on covariance analysis and theoretical calculations, section 5 takes care of the results and discussion, and section 6 concludes the manuscript.

%--------------------------------------

\section{Experimental Details}
\label{sec: Experimental details}

The experiment was carried out at Variable Energy Cyclotron Center (VECC), Kolkata, India using the K-130 cyclotron \cite{cy1}. In this experiment, helium was used to generate the alpha particles using the penning ionization gauge (PIG) ion source.\\ 
We have used the stacked foil activation technique \cite{17,18,19,20} followed by the off-line $\gamma$-ray spectroscopy to determine the nuclear reaction cross-sections of alpha-induced reactions on $^{nat}$Zn in the energy range from the threshold energy of reactions up to 37 MeV. In the stacked foil activation method, a particle beam was used to irradiate a stack of multiple thin foils together with a monitor foil. A catcher foil is attached to each target foil to record recoiled radioactive products from the target foil. Al was used as a catcher foil, as it produces only short lived radioactive isotopes in particular energy range, and to reduce gamma attenuation it should has low Z-material.

%--------------------------------------
\begin{figure}[b]
\begin{center}
\includegraphics[scale=0.5]{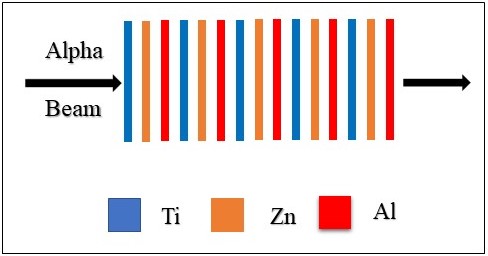}
\caption{ The schematic presentation of the monitor-target-catcher foil arrangement.}
\label{240pu}
\end{center}
\end{figure}
%------------------------------
In this experiment, $^{nat}$Zn (10 x 10 \textit{mm$^2$}), $^{nat}$Al (10 x 10 \textit{mm$^2$}) and  $^{nat}$Ti (10 x 10 \textit{mm$^2$}) thin metallic foils were used. The $^{nat}$Zn foil was used as target foil, while the $^{nat}$Al foil was used as catcher and energy degrader foil. The $^{nat}$Ti foil was used as monitor foil and the nuclear reaction   $^{nat}$Ti($\alpha$,x)$^{51}$Cr was used to calculate incident flux on the target foils.  The thickness of thin metallic foils of $^{nat}$Zn, $^{nat}$Al  and $^{nat}$Ti  were 7.85 $\pm$ 0.05 \textit{mg/cm$^2$}, 13.5 $\pm$ 0.08 \textit{mg/cm$^2$} and 1.80 $\pm$ 0.01 \textit{mg/cm$^2$} respectively. In this work, two different stacks were irradiated to cover the entire excitation function in the incident alpha energy range from the threshold energy up to 37 MeV. We attached an $^{nat}$Al foil to a $^{nat}$Zn foil in one set of stacks and each stack consisted of five such sets. In each stack, we placed one monitor foil ($^{nat}$Ti) before every target foil ($^{nat}$Zn). The details regarding irradiation time, incident energy of alpha particle and beam current for both stacks are given in Table 1. A systematic arrangement of the stacked foils is shown in Fig. 1. In the present experiment two collimators of diameter 8 mm were used and also a Faraday cup was used for current measurement placed  after the  samples. The charge particle beam from the cyclotron travels a considerable distance after the last quadruple magnet in the beamline before reaching the target foils. As a result of which, considerable defocusing in the beam can occur. The collimator prevents any stray beam, which does not fall on the target foils to reach the Faraday cup. The collimator thus helps in weeding out error in beam current measurement coming from such defocussed beam. We have used the Stopping and Range of Ions in Matter (SRIM-2008) code to determine the energy loss in a particular foil \cite{21,22}. This code provides us the information about energy loss of incident ions and range inside the matter.

%-----------------------------

\begin{table*}[htp]
\begin{center}
\caption{Irradiation condition and energy range for both stacks in the experiment.}
\label{tab: table 1}
\begin{tabular}{ccccccc}
\hline
Stack Number& ~~Incident Energy & ~~Energy Range  & ~~Irradiation Time & ~~Current \\
\vspace{2mm}
& ~~(MeV)& ~~(MeV)& ~~(hr) & ~~(nA)\\
\hline
\vspace{1mm}
Stack 1& ~~37& 37-22.2& 7& ~~~150\\
\vspace{1mm}
Stack 2& ~~32& 32-14.5& 7& ~~~150\\
\hline
\end{tabular}
\end{center}
\end{table*}
%--------------------------------------

%-------------------------------------
\section{Gamma-ray Spectrometry}
After irradiation, the activated samples were taken from the experiment hall to the counting room to detect the gamma-ray activity of the samples. After the target holder was safely opened, the monitor foils and the target foils were separated. For gamma-ray activity counting, both the target and the respective catcher foils were wrapped in small thin polythene bags, sealed to prevent from any contamination and placed on perspex plates. Depending on the half-life of the produced radionuclide, counting was started after various cooling intervals from the end of irradiation. In the present study, we have used a high purity germanium detector (HPGe) to detect the gamma-ray activity of the samples. The efficiency of the HPGe detector for different gamma energies was calculated by using a \textit{$^{152}$Eu} point source which has initial activity A$_0 = 3.908 \times10^{4}$ $\pm$ 197.68 Bq reported on 17 May 1982.  The standard \textit{$^{152}$Eu} point source has a half-life of T$_{1/2}$ = 13.517 $\pm$  0.009 years. The following equation is used to calculate the detection efficiency of the HPGe detector for a source-detector distance of 62.5 mm \cite{23}:
%--------------------------------------------
\begin{equation}
\varepsilon_{p} = \frac{CK_{c}}{A_{0}I_{\gamma}\Delta{t}e^{-{\lambda}t}} 
\end{equation}
%--------------------------------------------
In the above equation, $\epsilon_p$ represents the efficiency for the point source, $\lambda$ is the decay constant of \textit{$^{152}$Eu} point source,  A$_{0}$ represents the known activity of a standard \textit{$^{152}$Eu} point source, C denotes the total number of counts taken in $\Delta$t = 10000 seconds for ${\gamma}$-ray energy with absolute intensity ($I_\gamma$), K$_C$ denotes the summing correction factor, and t denotes the cooling time for the point source. A $\gamma$-ray spectrum of the irradiated target foil at incident alpha energy 36.32 MeV is shown in Fig. 2. The energy of the incident alpha beam is taken in the middle of every foil.
%------------------------------------------
\begin{figure}[]
\begin{center}
\includegraphics[scale=0.6]{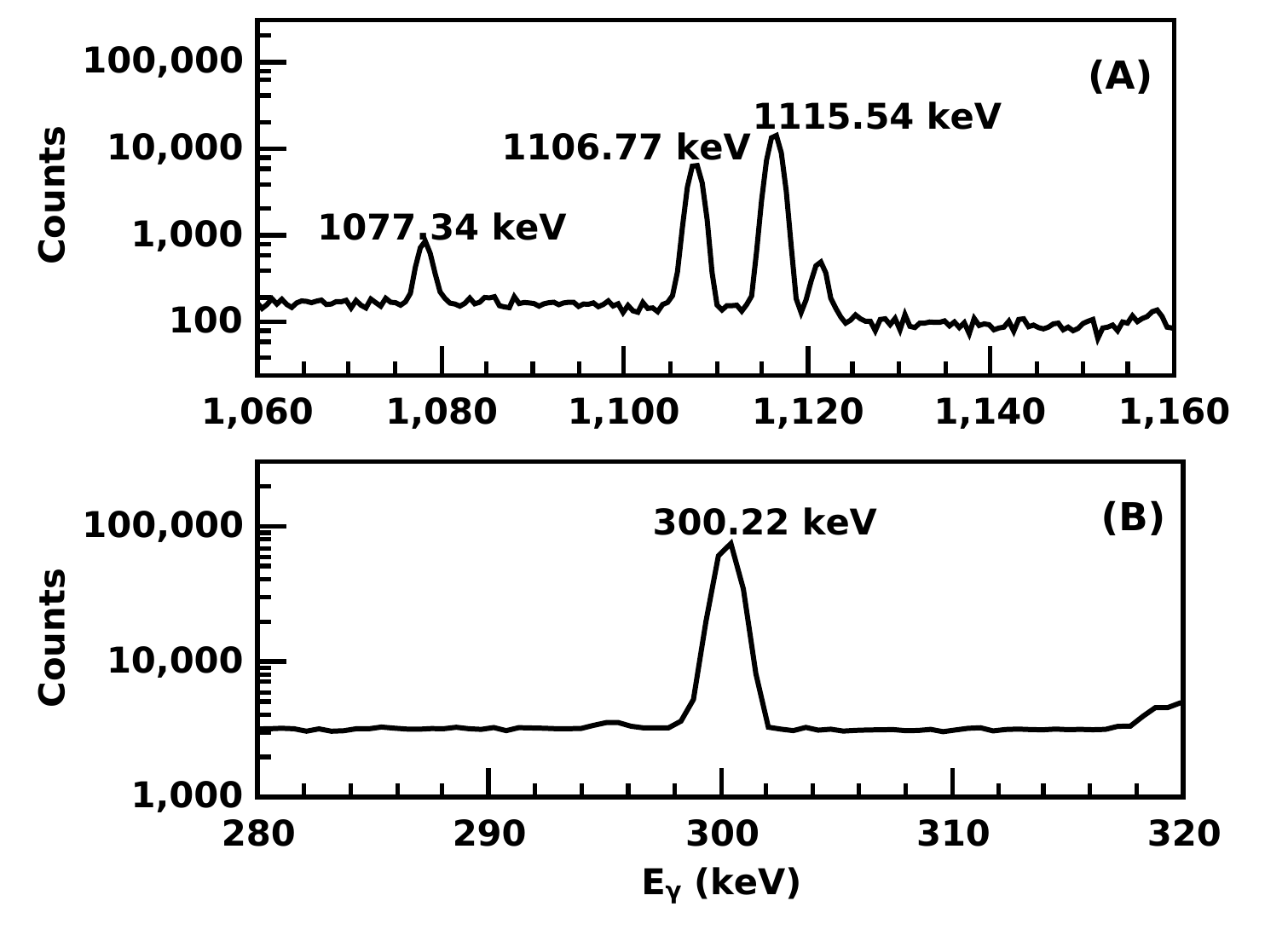}
\caption{ A $\gamma$-ray spectrum of the irradiated target foil at incident alpha energy 36.32 MeV. 
(a) for radionuclides $^{68}$Ge (1077.34 keV), $^{69}$Ge (1106.77 keV) and $^{65}$Zn (1115.54 keV) (b) for radionuclide $^{67}$Ga (300.22 keV)}
\label{240pu}
\end{center}
\end{figure}

%-------------------------------------

%-----------------------------------------
\begin{figure}[]
\begin{center}
\includegraphics[scale=0.6]{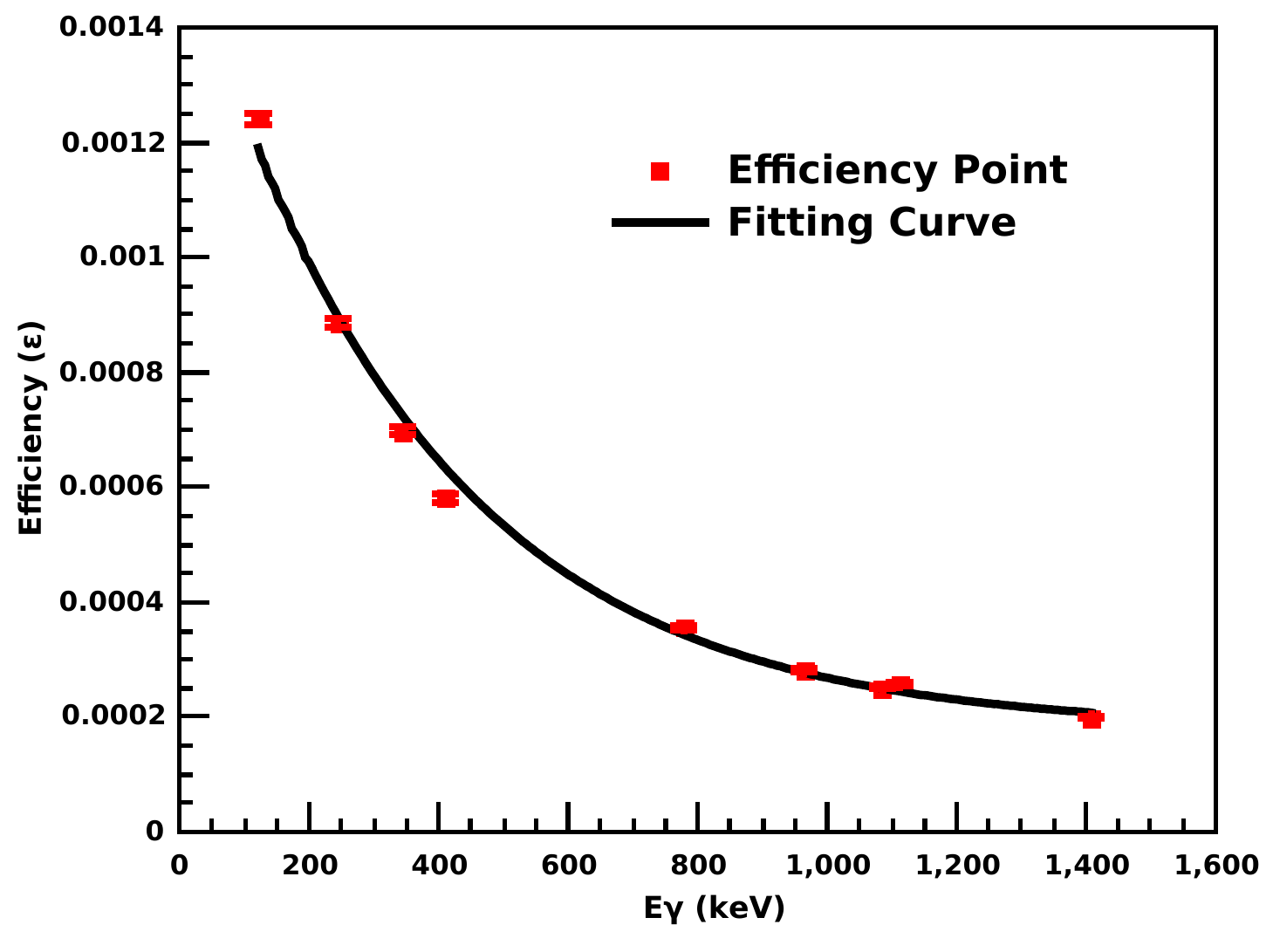}
\caption{ The HPGe detector efficiency curve for a distance of 62.5 mm between the detector and the source.}
\label{240pu}
\end{center}
\end{figure}
%--------------------------------------------
The  samples were of a finite area and the standard source \textit{$^{152}$Eu} was a point source, the efficiency of the point source geometry ($\epsilon_p$) has to be transferred  to the efficiency of the sample geometry ($\epsilon$). The Monte Carlo simulation code EFFTRAN  \cite{24,25} was used to transfer the efficiency from the point source geometry ($\epsilon_p$) to the sample geometry ($\epsilon$) and calculate the correction factor (K$_C$) of the coincidence summing effect. 
\\
The calculated efficiency value for sample source geometry ($\epsilon$) and point source geometry ($\epsilon_p$) placed at 62.5 mm from the detector are given in Table 3 with the correction factor (K$_C$). 
To calculate the efficiency of a particular $\gamma$-ray of the product radionuclide, we have used equation 2 which is a fitting function of interpolating the point-wise efficiencies $\epsilon$($E_{\gamma}$) of the $\gamma$-ray energy ($E_{\gamma}$) of the standard source \textit{$^{152}$Eu} \cite{23,26}.
%--------------------------------------------
\begin{equation}
\varepsilon(E_{\gamma}) = \varepsilon_{c} + \varepsilon_{o}exp({-E_{\gamma}/E_{0}})
\end{equation}
%--------------------------------------------
Here $\varepsilon_{c}$, $\varepsilon_{o}$ and $E_{0}$ are the detector efficiency ($\varepsilon$) fitting parameters. The value of these fitting parameters, uncertainties and their correlation matrix are given in Table 2. The efficiency curve of the HPGe detector is shown in Fig. 3.

%--------------------------------------
\begin{table*}[t]
\begin{center}
\caption{The value of the detector efficiency ($\varepsilon$) fitting parameters, as well as their uncertainties.} 
\begin{tabular}{ccccccc}
\hline\\
\vspace{3mm}
Parameters&Value& ~~~~~~~~~~~~~~~~Correlation matrix\\
\hline
\vspace{1mm}
$\varepsilon_{c}$ & 1.8$\times10^{-4}$ $\pm$  1.4$\times10^{-5}$& 1  \\
\vspace{1mm}
$\varepsilon_{0}$ & 1.4$\times10^{-3}$ $\pm$  7.4$\times10^{-5}$ &0.522&1\\
\vspace{1mm}
E$_{0}$ (keV) & 362 $\pm$ 28 &-0.892 & -0.767 & 1 \\
\hline
\end{tabular}
\end{center}
\end{table*}
%-------------------------------------
\begin{table*}[t]
\begin{center}
\caption{The HPGe detector's efficiency for both sample ($\varepsilon$) and point source ($\varepsilon_{p}$) geometries at different $\gamma$-ray energies with their $\gamma$-ray intensities and coincidence summing correction factor (K$_c$).} 
\begin{tabular}{cccccc}
\hline \\
\vspace{3mm}
$E_{\gamma}$(keV)&~$I_{\gamma}$ (\%)&~Counts(C)&~K$_{c}$&~$\varepsilon_{p}$  &~$\varepsilon$  \\

\hline 
\vspace{1mm} \\
121.78 &~28.53 $\pm$ 0.16&~ 376755 $\pm$ 640 &~1.02& ~0.00138 &~0.00124 $\pm$ 0.0000096 \\
\vspace{1mm}
244.69 &~7.55 $\pm$ 0.04 &~ 70512 $\pm$ 296 &~1.03 &~0.00098 &~0.00088 $\pm$ 0.0000075 \\
\vspace{1mm}
344.27 &~26.59 $\pm$ 0.2 &~ 197352 $\pm$ 454 &~1.02 &~0.00077 &~0.00069 $\pm$ 0.0000065 \\
\vspace{1mm}
411.11 &~2.24 $\pm$ 0.01 &~13571 $\pm$ 137 &~1.03 &~0.00064 &~0.00058 $\pm$ 0.0000074 \\
\vspace{1mm}
778.9 &~12.93 $\pm$ 0.08 &~48579 $\pm$ 288 &~1.02 &~0.00039 &~0.00035 $\pm$ 0.0000035\\
\vspace{1mm}
964.05 &~14.51 $\pm$ 0.07 &~43529 $\pm$ 231 &~1.01 &~0.00031 &~0.00028 $\pm$ 0.0000025 \\
\vspace{1mm}
1085.83 &~10.11 $\pm$ 0.05 &~27617 $\pm$ 177 &~0.99 &~0.00028 &~0.00025 $\pm$ 0.0000024 \\
\vspace{1mm}
1112.94 &~13.67 $\pm$ 0.08 &~37596 $\pm$ 203 &~1.0 &~0.00028 &~0.00026 $\pm$ 0.0000024 \\
\vspace{1mm}
1408.01 &~20.87 $\pm$ 0.09 &~44423 $\pm$ 218 &~1.0 &~0.00022 &~0.00019 $\pm$ 0.0000016 \\
\hline
\end{tabular}
\end{center}
\end{table*}

%----------------------------
\begin{table*}[htp]
 \centering
 \begin{tiny}

\caption{Nuclear reactions and details about other parameters of the radionuclides generated through $^{nat}$Zn($\alpha$,x) reactions.} 
\begin{tabular}{ccccccccc}
\hline\\

Radionuclide&Half-life& Decay mode&$E_{\gamma }$& $I_{\gamma }$&Reaction& Q-value\\
\vspace{2mm}
&($t_{1/2}$)&(\%)&(keV)&(\%)&&(MeV)\\
\hline
\vspace{1mm}\\
$^{65}$Zn &243.93 $\pm$ 0.09 days &$ec + \beta^+ (100)$ &1115.54 &50.04 $\pm$ 0.1 &$^{64}$Zn($\alpha, He^3$) & -12.59 \\
&&&&& $^{64}$Zn($\alpha$, n+2p) &-20.31\\
&&&&& $^{64}$Zn($\alpha$, p+d) &-18.09\\
&&&&&$^{66}$Zn($\alpha$, a+n) &-11.05\\
 
\vspace{1mm}
$^{67}$Ga &3.2617 $\pm$ 0.0005 days&$ec~(100)$ &300.22  & 16.64 $\pm$ 0.12 & $^{64}$Zn($\alpha$,p) & -3.99\\
&&&&  & $^{66}$Zn($\alpha$,t) & -14.55\\
&&&&& $^{66}$Zn($\alpha$,2n+p) &-23.03\\
&&&&&$^{67}$Zn($\alpha$,3n+p) &-30.08\\

\vspace{1mm}
$^{68}$Ge & 270.93 $\pm$ 0.13 days&$ec~(100)$ & - & - & $^{64}$Zn($\alpha$,$\gamma$) & 3.39\\
&& & && $^{66}$Zn($\alpha$,2n) & -15.64\\
&& & && $^{67}$Zn($\alpha$,3n) & -22.69\\
&& && & $^{68}$Zn($\alpha$,4n) & -32.88\\
\vspace{1mm}
$^{68}$Ga & 67.71 $\pm$ 0.08 min & $ec + \beta^+ (100)$ & 1077.34&3.22 $\pm$ 0.03 & $^{68}$Ge (ec) & \\

\vspace{1mm}
$^{69}$Ge & 39.05 $\pm$ 0.1 h & ~$ec + \beta^+ (100)$ & 1106.77&36.0 $\pm$ 0.4 & $^{66}$Zn($\alpha$,n) & -7.44\\
&& && & $^{67}$Zn($\alpha$,2n) & -14.49\\
&&&& & $^{68}$Zn($\alpha$,3n)  & -24.69\\
\hline
\end{tabular}

\end{tiny}
\end{table*}
%--------------------------------------

%--------------------------------------
\section{DATA ANALYSIS}
\subsection{Estimation of the reaction cross section}
In the present study, The following activation formula was used to calculate the cross-sections of the nuclear reactions;

\begin{equation}
\sigma_s= \sigma_m \frac{ \lambda_s C_s {I_m} \varepsilon_m N_m(1-e^{-\lambda_m t_{im}})(e^{-\lambda_m t_{cm}})(1-e^{-\lambda_m t_{am}})}{\lambda_m C_m {I_s} \varepsilon_s N_s(1-e^{-\lambda_s t_{is}})(e^{-\lambda_s t_{cm}})(1-e^{-\lambda_s t_{as}})}
\end{equation}

Here $\sigma_s$, $\sigma_m$ are the cross-sections of the sample and monitor nuclear reactions, $\lambda_s$, $\lambda_m$ are the decay constants for the sample and monitor nuclear reactions, $C_s$, $C_m$ are the peak area counts for the sample and monitor foils, $I_s$, $I_m$ represent the gamma-ray intensities of the produced radioisotopes from the sample and monitor foils, $\varepsilon_s$, $\varepsilon_m$ are the detector efficiencies for the sample and monitor nuclear reactions and $N_s$, $N_m$ are the particle number densities for the sample and monitor foils. In equation 3, $(t_i)_{s,m}$, $(t_c)_{s,m}$ and $(t_a)_{s,m}$ are the irradiation time, cooling time and counting time for the sample and monitor foils respectively. \\ To determine the uncertainty of the measured nuclear reaction cross-section, the uncertainty in the parameters contributing to the cross-section is taken into account such as monitor cross-section, detector efficiency, gamma-ray intensity, the particle number density in target, peak area counts and decay constant. The details regarding nuclear reactions, half-life, decay data and Q-value of the reactions are given in the Table 4.

\subsection{Covariance Analysis }

The cross-correlation between several measured values can be used to explain the detailed uncertainty in covariance analysis. The covariance matrix (I$_\sigma$) of the cross-section can be represented as \cite{28,29};

\begin{equation}
I_\sigma = S_xC_xS_x^T
\end{equation}

In the above equation,  I$_\sigma$ represents the covariance matrix of the measured nuclear reaction cross-sections of order r $\times$ r and $C_x$ matrix of order m $\times$ m, represents the semi covariance matrix of different variables in the cross-section formula  (equation 3) i.e. peak area counts ($C_\gamma$),  detector efficiency $\varepsilon(E_\gamma)$, flux ($\phi$), decay constant ($\lambda$), $\gamma$-ray intensity ($I_\gamma$), particle number density in the target ($N_t$). Here, $S_x$ represents the sensitivity matrix with the corresponding element (j, k);

\begin{equation}
S_{xjk} = \frac{\partial \sigma_j}{\partial x_k}; (j = 1,2,3,...r;~k = 1,2,3,...m) 
\end{equation}
Here, the total number of the measured cross-sections for a nuclear reaction is equal to r and the total number of variables in the cross-section formula is equal to m. If two variables $x_{k}$ and $x_{l}$ (k, $l$ = 1, 2, 3,.....m) are required in the calculation of the cross-sections, then we can write covariance matrix (C$_x$) of these variables as follows \cite{30};

\begin{equation}
C_x(x_k,x_l) = Cor(x_k,x_l)(\Delta x_k\Delta x_l) 
\end{equation}
In the above equation, the term $Cor(x_k,x_l)$ represents the correlation coefficient between two attributes $x_k$, $x_l$ and it has a value in the range of 0 to 1. If k=$l$ then the value of the term $Cor(x_k,x_l)$ is equal to 1, in which case these two variables $x_k$, $x_l$ are fully correlated. The interpolated efficiency, error and correlation matrix of the $\gamma$-ray for the nuclear reactions of the sample and monitor are given in Table 5.

The percentage uncertainties of the different parameters contributing to the uncertainties of the sample nuclear reaction cross-sections are given in Table 6. The calculated spread in the incident alpha beam energy for each energy point is given in Tables 7-10 and shown in Figs. 4-7.

%------------------------------------------

\begin{table*}
\centering
\begin{scriptsize}
\caption{ The Interpolated efficiency, error and correlation matrix of the $\gamma$-ray for the nuclear reactions of the sample and monitor.}

\begin{tabular}{cccccccccccc}
\hline
\hline \\
Reaction&E$_\gamma$ (keV)&Efficiency ($\epsilon$) &Error ($\Delta \epsilon$)&~~~~~~~~~~{Correlation matrix}\\
\\

\hline\\
\vspace{1mm}
$^{nat}$Ti($\alpha$,x)$^{51}$Cr&320.08 &0.00076&0.0000176& 1 \\
\vspace{1mm}
$^{nat}$Zn($\alpha$,x)$^{65}$Zn&1115.50 &0.00024&0.0000055
 &0.0953&1  &\\

\hline\\
\vspace{1mm}
$^{nat}$Ti($\alpha$,x)$^{51}$Cr&320.08 &0.00076&0.0000176 & 1 \\
\vspace{1mm}
$^{nat}$Zn($\alpha$,x)$^{68}$Ge&1077.30 &0.00025&0.0000056
 &0.174&1  &\\
\hline

\\
\vspace{1mm}
$^{nat}$Ti($\alpha$,x)$^{51}$Cr&320.08 &0.00076&0.0000176 & 1 \\
\vspace{1mm}
$^{nat}$Zn($\alpha$,x)$^{69}$Ge&1106.77 &0.00024&0.0000055
 &0.114&1  &\\

\hline\\
\vspace{1mm}
$^{nat}$Ti($\alpha$,x)$^{51}$Cr&320.08 &0.00076&0.0000176 & 1 \\
\vspace{1mm}
$^{nat}$Zn($\alpha$,x)$^{67}$Ga&300.22 &0.00079&0.0000182
 &0.994&1  &\\
\hline
\hline
\end{tabular}
\end{scriptsize}

\end{table*}

%----------------------------------------------
\begin{table*}
\begin{scriptsize}
\centering
\caption{ The percentage uncertainties of the different parameters contributing to the uncertainties of the sample nuclear reaction cross-sections}
\begin{tabular}{ccccccc} 
\hline
\\

Parameters &~~~~~~~$^{nat}$Zn($\alpha$,x)$^{65}$Zn&~~~~~~~ $^{nat}$Zn($\alpha$,x)$^{68}$&~~~~~~~ $^{nat}$Zn($\alpha$,x)$^{69}$Ge&~~~~~~~ $^{nat}$Zn($\alpha$,x)$^{67}$Ga\\
\hline\\
$x_i$& ~~~~~~~$\Delta x_i~(\%)$& ~~~~~~~$\Delta x_i~(\%)$& ~~~~~~~$\Delta x_i~(\%)$& ~~~~~~~$\Delta x_i~(\%)$\\

\\
\hline
\vspace{1mm}
$\sigma_m$& ~~~~~~~4-6& ~~~~~~~4-6& ~~~~~~~4-7& ~~~~~~~4-7&  \\
\vspace{1mm}
$C_s$& ~~~~~~~0.5-5& ~~~~~~~0.5-4& ~~~~~~~0.5-2& ~~~~~~~0.5-1.5&  \\
\vspace{1mm}
$C_m$& ~~~~~~~ 0.5-1& ~~~~~~~0.5-1& ~~~~~~~0.5-1& ~~~~~~~0.5-1&  \\
\vspace{1mm}
$\lambda_m$& ~~~~~~~0.01& ~~~~~~~0.01& ~~~~~~~0.01& ~~~~~~~0.01&  \\

\vspace{1mm}
$\lambda_s$& ~~~~~~~0.04& ~~~~~~~0.05& ~~~~~~~0.26& ~~~~~~~0.02&  \\

\vspace{1mm}
$I_m$& ~~~~~~~0.1& ~~~~~~~0.1& ~~~~~~~0.1& ~~~~~~~0.1&  \\

\vspace{1mm}
$I_S$& ~~~~~~~0.2& ~~~~~~~0.93& ~~~~~~~1.11& ~~~~~~~0.72&  \\

\vspace{1mm}
$N_m$& ~~~~~~~0.56& ~~~~~~~0.56& ~~~~~~~0.56& ~~~~~~~0.56&  \\

\vspace{1mm}
$N_s$& ~~~~~~~0.67& ~~~~~~~0.67& ~~~~~~~0.67& ~~~~~~~0.67&  \\

\vspace{1mm}
$\epsilon_m$& ~~~~~~~2.31& ~~~~~~~2.31& ~~~~~~~2.31& ~~~~~~~2.31&  \\
$\epsilon_s$& ~~~~~~~2.25& ~~~~~~~2.24& ~~~~~~~2.25& ~~~~~~~2.30&  \\
\hline
\end{tabular}
\end{scriptsize}
\end{table*}

%--------------------------------------

%-------------------------------------
\subsection{Theoretical Calculations}
We have used the statistical nuclear model code TALYS-1.9 \cite{15} for the theoretical calculations of the reactions $^{nat}$Zn($\alpha$,x)$^{68}$Ge, $^{nat}$Zn($\alpha$,x)$^{69}$Ge, \\$^{nat}$Zn($\alpha$,x)$^{65}$Zn and  $^{nat}$Zn($\alpha$,x)$^{67}$Ga. The TALYS is a Fortran-based nuclear reaction model code which is used to calculate different physical observables related to nuclear reactions. This nuclear code is based on the Hauser–Fesbach statistical model and it contains distinct choices for level density and optical model parameters \cite{31}. In this nuclear code, we can do calculations for nuclear reactions having projectiles such as photons, neutrons, protons, tritons, deuterons,  $^3$He- and alpha-particles and target nuclides with masses of 12 and larger in the 1 keV - 200 MeV energy range. 
The TALYS have six different level density models. The ldmodel-1 is related to the constant temperature and the Fermi gas model, ldmodel-2 is related to the back-shifted Fermi gas model, ldmodel-3 is related to the generalized superfluid model, ldmodel-4 is from the Gorley table (Skyrme Force), ldmodel-5 is  from Hilaire’s combinatorial tables (Skyrem force) and ldmodel-6 is from Hilaire’s combinatorial tables (temperature-dependent HFB, Gogni force). Among these six level density models ldmodel-1, 2, 3 are phenomenological level density models and ldmodel-4, 5, 6 are microscopic level density models \cite{35,36,37,38,39,40}. In the present work we have used all these six level density models and the results of the theoretical calculations were compared with the experimentally obtained nuclear reaction cross-sections. 

%-----------------------
\section{RESULTS AND DISCUSSION}

We have reported reaction cross-sections, uncertainties and a covariance matrix of $^{nat}$Zn($\alpha$,x) nuclear reactions for the projectile energy range from the corresponding threshold energy for each contributing reaction up to 37 MeV. In the present work, the measured nuclear reaction cross-sections compared to the theoretical prediction from the TALYS nuclear reaction code, evaluated data from TENDL-2019 and the existing experimental data from EXFOR.
The excitation functions of nuclear reactions are shown in Figs. 4-7 and measured reaction cross-sections with their correlation matrices are presented in Tables 7-10. 
%-----------------------------
\subsection{Production cross-section of $^{65}$Zn}
The measured nuclear reaction cross-section value for the $^{nat}$Zn($\alpha$,x)$^{65}$Zn nuclear reaction is presented in Fig. 4 along with the theoretical excitation function from the TALYS code, evaluated nuclear data from TENDL-2019 and previously calculated cross-sections available on the EXFOR. The cross-sections for the $^{nat}$Zn($\alpha$,x)$^{65}$Zn  nuclear reaction were estimated using a $\gamma$-ray with an energy of 1115.54 keV and intensity of 50.04 $\%$ that arises from the decay of $^{65}$Zn radionuclide. The calculated experimental results for $^{nat}$Zn($\alpha$,x)$^{65}$Zn reaction are in good agreement with the existing reaction data given by A. Karpeles and Y. Nagame \textit{et al.} \cite{41,42}, as shown in Fig. 4. The theoretical results from ldmodel-2 (represented in red colour by a solid line) follow the trend of excitation function of this nuclear reaction. There is a good agreement between evaluated nuclear data from TENDL-2019 and the theoretical result from ldmodel-1 for the nuclear reaction $^{nat}$Zn($\alpha$,x)$^{65}$Zn. 
The obtained reaction cross-sections, as well as their uncertainties and correlation matrix for the reaction $^{nat}$Zn($\alpha$,x)$^{65}$Zn are given in Table 7.
%---------------------
\begin{figure}[]
\begin{center}
\includegraphics[scale=0.6]{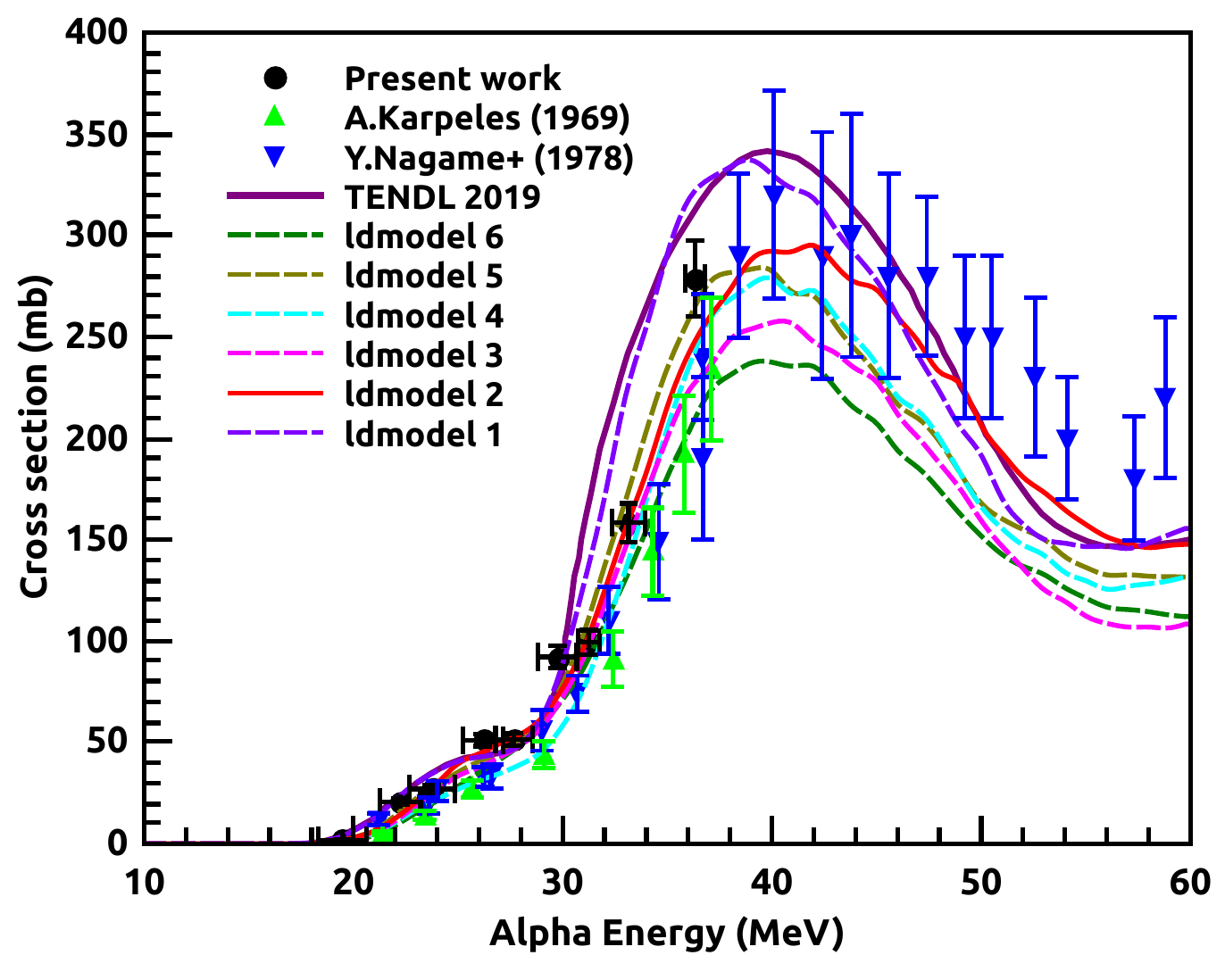}
\caption{Cross sections for $^{nat}$Zn($\alpha$,x)$^{65}$Zn reaction from this study in comparison of the available experimental data from EXFOR and theoretical calculation from TALYS.}
\label{240pu}
\end{center}
\end{figure}
%-----------------------
\begin{figure}[]
\begin{center}
\includegraphics[scale=0.6]{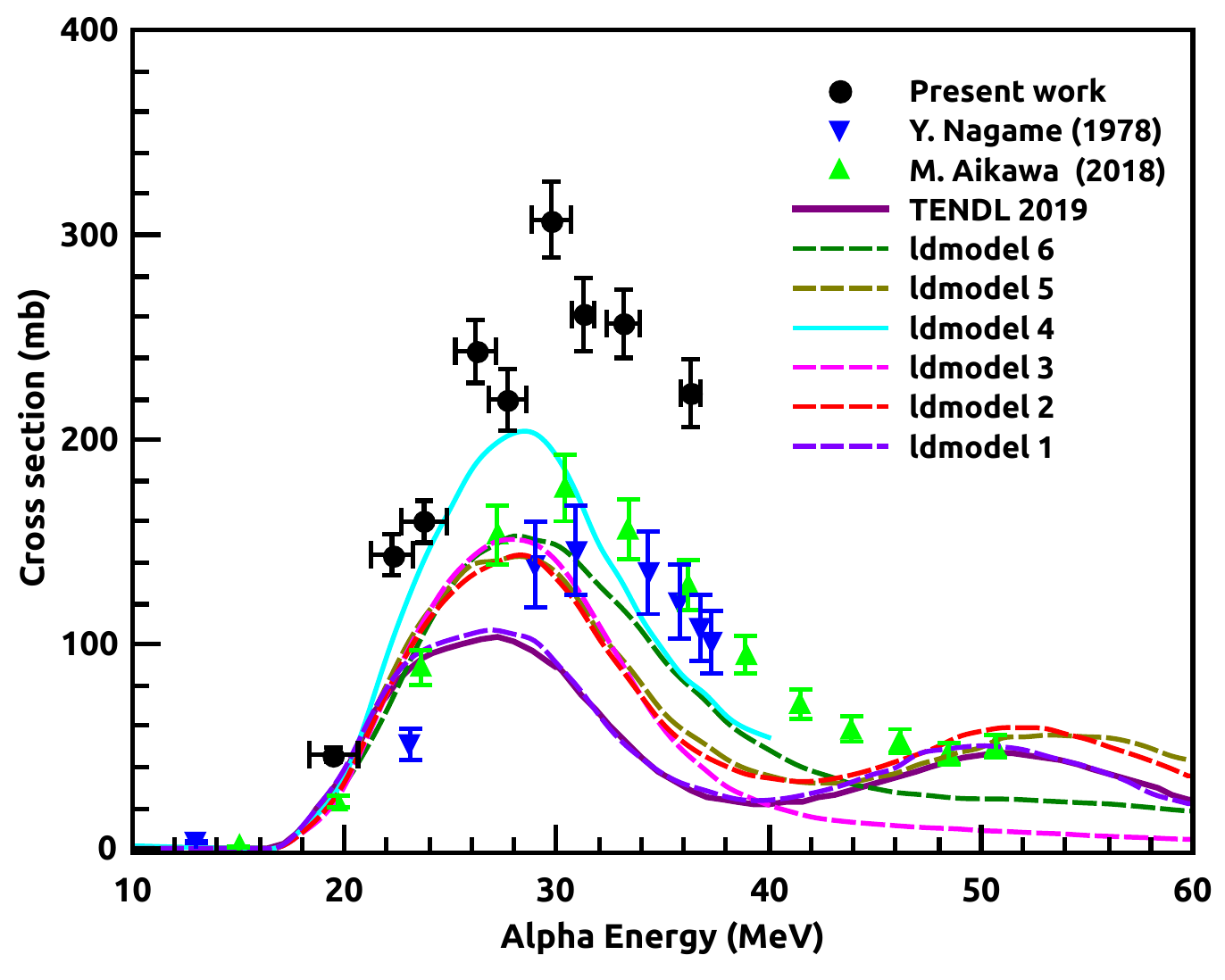}
\caption{Cross sections for $^{nat}$Zn($\alpha$,x)$^{68}$Ge reaction from this study in comparison of the available experimental data from EXFOR and theoretical calculation from TALYS.}
\label{240pu}
\end{center}
\end{figure}
%--------------------------------------
\begin{table*}[!htbp]
\centering
\begin{scriptsize}
\caption{The calculated reaction cross-section, uncertainty and correlation matrix of the nuclear reaction $^{nat}$Zn($\alpha$,x)$^{65}$Zn.}

\begin{tabular}{cccccccccccc}
\hline \\
E$_{\alpha}$ (MeV)&Cross-section (mb)&\multicolumn{4}{c}~~~~~~~~~~~~~{Correlation matrix}\\
&($\sigma$ $\pm$ $\bigtriangleup\sigma$)\\&&\\
\hline\\
\vspace{1mm}
19.47 $\pm$ 1.15&1.78 $\pm$ 0.14& 1 \\
\vspace{1mm}
22.23 $\pm$ 0.98&20.74 $\pm$ 1.41&0.193&1  &\\
\vspace{1mm}
23.75 $\pm$ 1.09&27.01 $\pm$ 1.78&0.199&0.228&1 &\\
\vspace{1mm}
26.18 $\pm$ 0.96&51.10 $\pm$ 2.99&0.224&0.257&0.265&1\\
\vspace{1mm}
27.68 $\pm$ 0.89&51.49 $\pm$ 3.0&0.225&0.258&0.267&0.300&1\\
\vspace{1mm}
29.75 $\pm$ 0.92&92.09 $\pm$ 5.44&0.222&0.255&0.263&0.296&0.297&1\\
\vspace{1mm}
31.25 $\pm$ 0.52&99.27 $\pm$ 6.09&0.214&0.245&0.253&0.285&0.286&0.283&1\\
\vspace{1mm}
33.15 $\pm$ 0.78&158.20 $\pm$ 9.63&0.216&0.247&0.255&0.287&0.288&0.285&0.274&1\\
\vspace{1mm}
36.32 $\pm$ 0.46&~278.78 $\pm$ 18.68&0.196&0.224&0.232&0.261&0.262&0.259&0.249&0.251&1\\
\hline
\end{tabular}

\end{scriptsize}
\end{table*}

%---------------------------------
%-----------------------

\begin{table*}[]
\centering
\begin{scriptsize}
    
\caption{The calculated reaction cross-section, uncertainty and correlation matrix of the nuclear reaction $^{nat}$Zn($\alpha$,x)$^{68}$Ge.} 

\begin{tabular}{ccccccccccccc}
\hline
\hline \\
E$_{\alpha}$ (MeV)&Cross-section (mb)&\multicolumn{4}{c}~~~~~~~~~~~~~{Correlation matrix}\\
&($\sigma$ $\pm$ $\bigtriangleup\sigma$)\\&&\\
\hline\\
\vspace{1mm}
19.47 $\pm$ 1.15 & 34.12 $\pm$ 2.49 & 1 \\
\vspace{1mm}
22.23 $\pm$ 0.98 &122.77 $\pm$ 8.31 &0.206 & 1  &\\
\vspace{1mm}
23.75 $\pm$ 1.09 & 137.44 $\pm$ 8.61 & 0.222 & 0.240 & 1 &\\
\vspace{1mm}
26.18 $\pm$ 0.96 & 219.70 $\pm$ 13.51 & 0.227 & 0.245 & 0.264 & 1\\
\vspace{1mm}
27.68 $\pm$ 0.89 & 206.37 $\pm$ 13.82 & 0.208 & 0.225 & 0.243 & 0.247 & 1\\
\vspace{1mm}
29.75 $\pm$ 0.92 & 294.34 $\pm$ 17.21 & 0.238 & 0.257 & 0.278 & 0.283 & 0.260 & 1\\
\vspace{1mm}
31.25 $\pm$ 0.52 & 260.66 $\pm$ 17.56 & 0.207 & 0.223 & 0.241 & 0.246 & 0.226 & 0.258 & 1\\
\vspace{1mm}
33.15 $\pm$ 0.78 & 256.54 $\pm$ 16.37 & 0.218 & 0.236 & 0.255 & 0.259 & 0.238 & 0.273 & 0.237 & 1\\
\vspace{1mm}
36.32 $\pm$ 0.46 & 222.56 $\pm$ 16.25 & 0.191 & 0.206 & 0.222 & 0.226 & 0.208 & 0.238 & 0.207 & 0.218 & 1\\
\hline
\hline
\end{tabular}
\end{scriptsize}
\end{table*}
%---------------------------------
\begin{table*}[]
\centering
\begin{scriptsize}
\caption{The calculated reaction cross-section, uncertainty and correlation matrix of the nuclear reaction $^{nat}$Zn($\alpha$,x)$^{69}$Ge.}
\vspace{2mm}
%---------------------------------------
\begin{tabular}{ccccccccccccc}
\hline
\hline\\

E$_{\alpha}$ (MeV)&Cross-section (mb)&\multicolumn{4}{c}~~~~~{Correlation matrix}\\
&($\sigma$ $\pm$ $\bigtriangleup\sigma$)\\&&\\

\hline\\
\vspace{1mm}
14.47 $\pm$ 1.17 & 232.55 $\pm$ 17.26 & 1 \\
\vspace{1mm}
19.48 $\pm$ 1.15 & 257.65 $\pm$ 16.64 & 0.235 & ~1& \\
\vspace{1mm}
22.23 $\pm$ 0.98 & 291.90 $\pm$ 17.77 &0.250 & 0.288  & 1\\
\vspace{1mm}
23.75 $\pm$ 1.09 & 171.25 $\pm$ 9.98 & 0.261 & 0.300 & 0.319 & 1&\\
\vspace{1mm}
26.18 $\pm$ 0.96 & 119.37 $\pm$ 6.94 & 0.262 & 0.301 & 0.320 & 0.334 & 1&\\
\vspace{1mm}
27.68 $\pm$ 0.89 & 85.28 $\pm$ 5.03 & 0.258 & 0.297 & 0.315 & 0.329 & 0.330 & 1&\\
\vspace{1mm}
29.75 $\pm$ 0.92 & 82.20 $\pm$ 4.91 & 0.254 & 0.293 & 0.311 & 0.325 & 0.326 & 0.321 & 1&\\
\vspace{1mm}
31.25 $\pm$ 0.52 & 75.20 $\pm$ 4.67 & 0.245 & 0.281 & 0.299 & 0.312 & 0.313 & 0.308 & 0.304 & 1&\\
\vspace{1mm}
33.15 $\pm$ 0.78 & 77.23 $\pm$ 4.79 & 0.245 & 0.282 & 0.300 & 0.313 & 0.314 & 0.309 & 0.306 & 0.294 & 1&\\
\vspace{1mm}
36.32 $\pm$ 0.46 & 105.50 $\pm$ 7.17 & 0.224 & 0.258 & 0.274 & 0.286 & 0.287 & 0.282 & 0.279 & 0.268 & 0.269 & 1&\\
\hline
\hline
\end{tabular}
\end{scriptsize}
\end{table*}

%---------------------------------
\subsection{Production cross-section of $^{68}$Ge}
In the present work, the measured nuclear reaction cross-section value for the $^{nat}$Zn($\alpha$,x)$^{68}$Ge nuclear reaction is presented in Fig. 5 along with the theoretical excitation function from the TALYS code, evaluated nuclear data from TENDL-2019 and previously calculated cross-sections available on the EXFOR. The cross-sections for the \\$^{nat}$Zn($\alpha$,x)$^{68}$Ge  nuclear reaction were estimated using a $\gamma$-ray with an energy of 1077.34 keV and intensity of 3.22 $\%$ that arises from the decay of $^{68}$Ga radionuclide. The radioisotope $^{68}$Ge decays to $^{68}$Ga with 100 $\%$ electron capture. The measured experimental results for the $^{nat}$Zn($\alpha$,x)$^{68}$Ge reaction are  higher than the existing experimental data given by Y. Nagame \textit{et al.} and M. Alkawa \textit{et al.} \cite{42,43}, as shown in Fig. 5. 
The theoretical results from ldmodel-4 (represented in cyan colour by a solid line) partially follow the trend of present experimental data. The theoretical results from ldmodel-4 are in good agreement with measured experimental results for this reaction in the energy range 19-28 MeV and results from the ldmodel-4 are lower then measured experimental results in the energy range 29-37 MeV. It is clear from Fig. 5 that the data evaluated by TENDL-2019 do not follow the experimental data obtained by us as well as the data reported by Y. Nagame \textit{et al.}  and M. Alkawa \textit{et al.} There is a good agreement between evaluated nuclear data from TENDL-2019 and the theoretical result from ldmodel-1 for the nuclear reaction $^{nat}$Zn($\alpha$,x)$^{68}$Ge. The obtained reaction cross-sections, as well as their uncertainties and correlation matrix for the nuclear reaction $^{nat}$Zn($\alpha$,x)$^{68}$Ge are given in Table 8.

\subsection{Production cross-section of $^{69}$Ge}
\begin{figure}[]
\begin{center}
\includegraphics[scale=0.6]{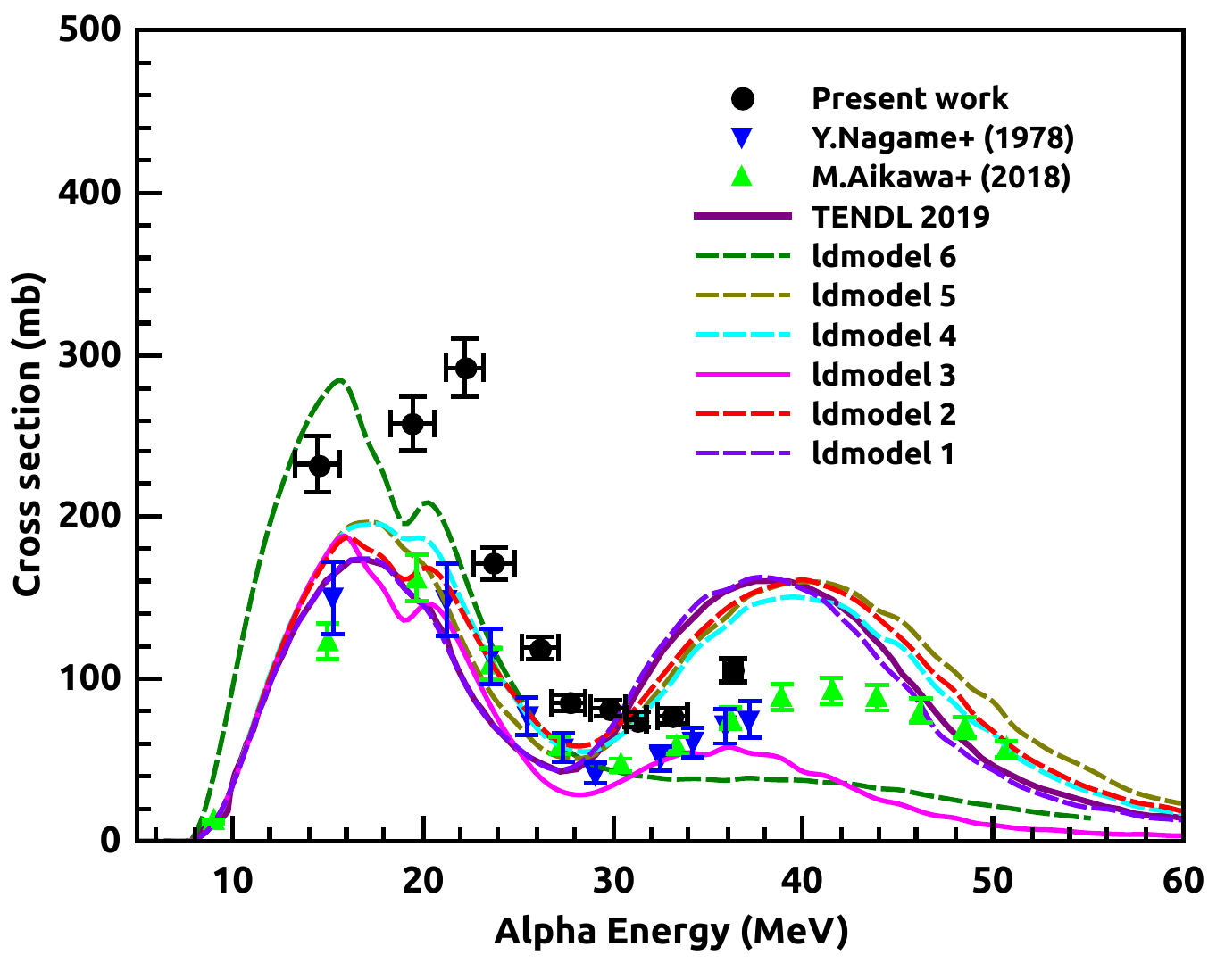}
\caption{Cross sections for $^{nat}$Zn($\alpha$,x)$^{69}$Ge reaction from this study in comparison of the available experimental data from EXFOR and theoretical calculation from TALYS.}
\label{240pu}
\end{center}
\end{figure}
%------------------------------------------

\begin{table*}[]
\centering
\begin{scriptsize}

\caption{The calculated reaction cross-section, uncertainty and correlation matrix of the nuclear reaction $^{nat}$Zn($\alpha$,x)$^{67}$Ga.}

\begin{tabular}{ccccccccccccc}
\hline
\hline \\
E$_{\alpha}$ (MeV)&Cross-section (mb)&\multicolumn{4}{c}~~~~~{Correlation matrix}\\
&($\sigma$ $\pm$ $\bigtriangleup\sigma$)\\&&\\
\hline\\
\vspace{1mm}
14.47 $\pm$ 1.17 & 486.98 $\pm$ 32.87 & 1 \\
\vspace{1mm}
19.48 $\pm$ 1.15 & 617.18 $\pm$ 34.74 & 0.036 & 1& \\
\vspace{1mm}
22.23 $\pm$ 0.98 & 536.43 $\pm$ 28.08 &0.038 & 0.046  & 1\\
\vspace{1mm}
23.75 $\pm$ 1.09 & 392.25 $\pm$ 21.78 & 0.036 & 0.044 & 0.047 & 1&\\
\vspace{1mm}
26.18 $\pm$ 0.96 & 214.52 $\pm$ 10.54 & 0.041 & 0.049 & 0.053 & ~0.050 & 1&\\
\vspace{1mm}
27.68 $\pm$ 0.89&168.23 $\pm$ 12.57 & 0.027& 0.032 & 0.035 & 0.033 & 0.037 & 1&\\
\vspace{1mm}
29.75 $\pm$ 0.92 & 97.09 $\pm$ 6.47 & 0.030 & 0.036 & 0.039 & 0.037 & 0.042 & 0.027 & 1&\\
\vspace{1mm}
31.25 $\pm$ 0.52 & 95.08 $\pm$ 8.99 & 0.021 & 0.026 & 0.027 & 0.026 & 0.029 & 0.019 & 0.022 & 1&\\
\vspace{1mm}
33.15 $\pm$ 0.78 & 67.37 $\pm$ 6.54 & 0.021 & 0.025 & 0.027 & 0.025 & 0.029 & 0.019 & 0.021 & 0.015 & 1&\\
\vspace{1mm}
36.32 $\pm$ 0.46 & 101.06 $\pm$ 7.29 & 0.028 & 0.034 & 0.036 & 0.034 & 0.038 & 0.025 & 0.028& 0.020 & 0.019 & 1&\\
\hline
\hline
\end{tabular}
\end{scriptsize}

\end{table*}

In the present work, the measured nuclear reaction cross-section value for the $^{nat}$Zn($\alpha$,x)$^{69}$Ge nuclear reaction is presented in Fig. 6 along with the theoretical excitation function from the TALYS code and previously calculated cross-sections available on the EXFOR. The cross-sections for the $^{nat}$Zn($\alpha$,x)$^{69}$Ge  nuclear reaction were estimated using a $\gamma$-ray with an energy of 1106.77 keV and intensity of 36 $\%$ that arises from the decay of $^{69}$Ge radionuclide. The calculated experimental results for $^{nat}$Zn($\alpha$,x)$^{69}$Ge reaction partially follow the trend of existing experimental data given by Y. Nagame \textit{et al.}  and M. Aikawa \textit{et al.} \cite{42,43}, as shown in Fig. 6. The measured experimental results for this reaction are higher than the existing experimental data in the energy range 14-20 MeV and are in good agreement with the existing experimental data in the energy range 21-37 MeV.  The theoretical results from ldmodel-6 are in good agreement with the measured experimental results for this reaction. It is clear from Fig. 6 that the data evaluated by TENDL-2019 do not follow the experimental data obtained by us as well as the data reported by Y. Nagame \textit{et al.}  and M. Alkawa \textit{et al.} There is a good agreement between evaluated nuclear data from TENDL-2019 and the theoretical result from ldmodel-1 for the nuclear reaction $^{nat}$Zn($\alpha$,x)$^{69}$Ge. The obtained reaction cross-sections, as well as their uncertainties and correlation matrix for the nuclear reaction $^{nat}$Zn($\alpha$,x)$^{69}$Ge are given in Table 9.

\begin{figure}[]
\begin{center}
\includegraphics[scale=0.6]{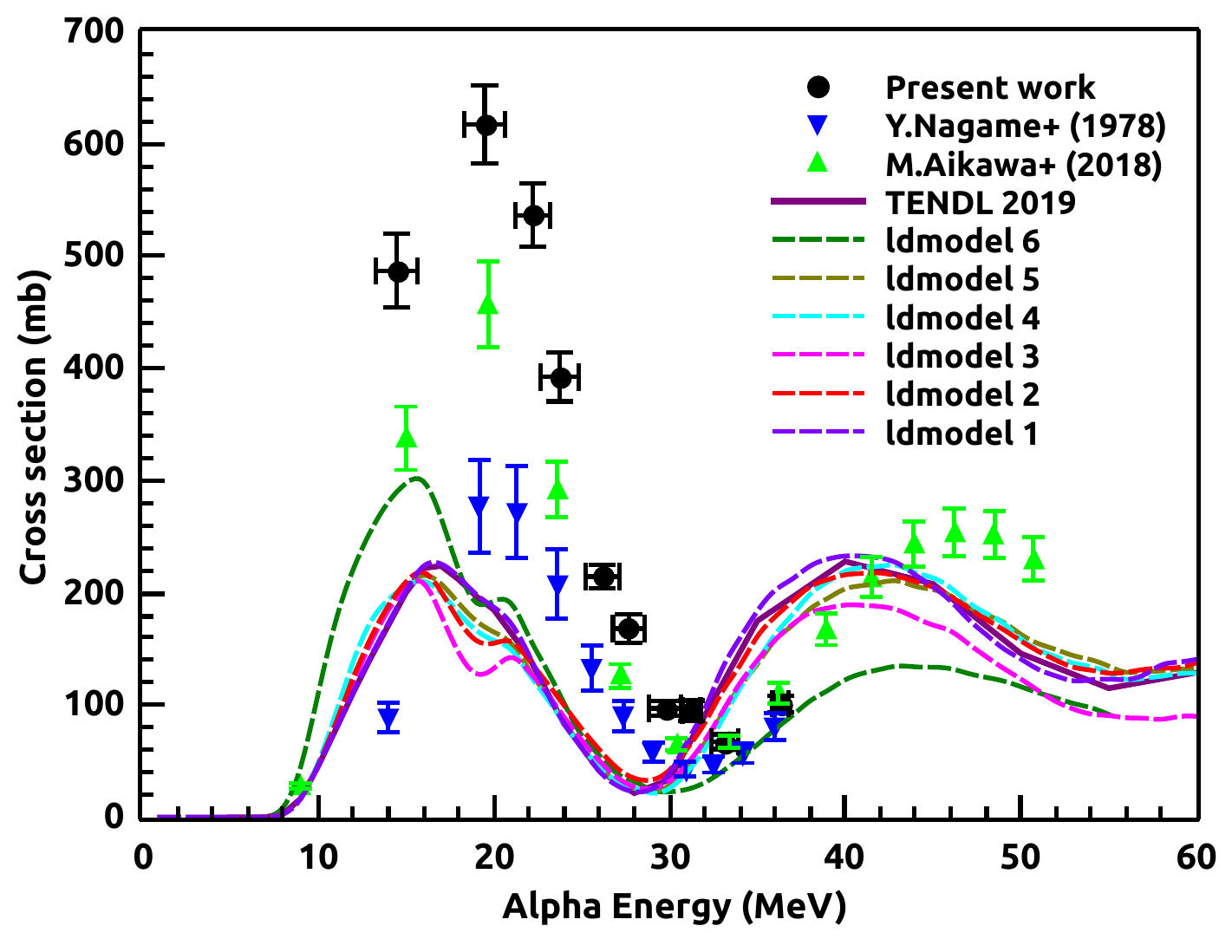}
\caption{ Cross sections for $^{nat}$Zn($\alpha$,x)$^{67}$Ga reaction from this study in comparison of the available experimental data from EXFOR and theoretical calculation from TALYS.}
\label{240pu}
\end{center}
\end{figure}
%-----------------------------------------

%--------------------------------
\subsection{Production cross-section of $^{67}$Ga}
In our work, the measured nuclear reaction cross-section value for the $^{nat}$Zn($\alpha$,x)$^{67}$Ga nuclear reaction is presented in Fig. 7 along with the theoretical excitation function from the TALYS code and previously calculated cross-sections available on the EXFOR. The cross-sections for the $^{nat}$Zn($\alpha$,x)$^{67}$Ga  nuclear reaction were estimated using a $\gamma$-ray with an energy of 300.22 keV and intensity of 16.64 $\%$ that arises from the decay of $^{67}$Ga radionuclide.\\ The measured experimental results for $^{nat}$Zn($\alpha$,x)$^{67}$Ga reaction partially follow the trend of existing experimental data given by Y. Nagame \textit{et al.}  and M. Aikawa \textit{et al.} \cite{42,43}, as shown in Fig. 7. The measured experimental results for this reaction are higher than the existing experimental data in the energy range 14-20 MeV and are in good agreement with the existing experimental data in the energy range 21-37 MeV. The theoretical results are slightly lower than the experimental results of this nuclear reaction in the energy range 10-30 MeV. It is clear from Fig. 7 that the data evaluated by TENDL-2019 do not follow the experimental data obtained by us as well as the data reported by Y. Nagame \textit{et al.}  and M. Aikawa \textit{et al.} There is a good agreement between evaluated nuclear data from TENDL-2019 and the theoretical result from ldmodel-1 for the nuclear reaction $^{nat}$Zn($\alpha$,x)$^{67}$Ga. The obtained reaction cross-sections, as well as their uncertainties and correlation matrix for the nuclear reaction $^{nat}$Zn($\alpha$,x)$^{67}$Ga are given in Table 10.
%--------------------------

%--------------------------------------
\section{Conclusion} 

In the present study, we have measured the cross-sections for $^{nat}$Zn($\alpha$,x)$^{68}$Ge, $^{nat}$Zn($\alpha$,x)$^{69}$Ge, $^{nat}$Zn($\alpha$,x)$^{65}$Zn and  $^{nat}$Zn($\alpha$,x)$^{67}$Ga nuclear reactions using the stack foil activation technique for the projectile energy range 14-37 MeV, a complete covariance analysis has also been performed. The detailed uncertainty analysis for above mentioned reactions, as well as their accompanying correlation matrix, is documented.\\
The measured  cross-sections for $^{nat}$Zn($\alpha$,x)$^{65}$Zn nuclear reaction are in good agreement with existing experimental data from the EXFOR, the measured cross-sections for $^{nat}$Zn($\alpha$,x)$^{68}$Ge, $^{nat}$Zn($\alpha$,x)$^{69}$Ge and $^{nat}$Zn($\alpha$,x)$^{67}$Ga nuclear reactions partially follow the existing experimental data. The ldmodel-4 provides the most accurate theoretical results for $^{nat}$Zn($\alpha$,x)$^{68}$Ge, ldmodel-6 provides the most accurate theoretical resutls for $^{nat}$Zn($\alpha$,x)$^{69}$Ge nuclear reaction and the ldmodel-2 provides the most accurate theoretical results for $^{nat}$Zn($\alpha$,x)$^{65}$Zn nucler reaction. The theoretical results of ldmodel-4 follow the trend of the excitation function above 30 MeV energy for the $^{nat}$Zn($\alpha$,x)$^{67}$Ga nuclear reaction. The ldmodel-2 is based on the back-shifted Fermi gas model, the ldmodel-4 based on Skyrme force from Goriely’s tables while the ldmodel-6 relates to Hilaire’s combinatorial tables
(temperature-dependent HFB, Gogni force). The evaluated data from the TENDL-2019 are consistent with the theoretical results obtained from the ldmodel-1. It looks from the present study that the evaluation results obtained from TENDL-2019 need corrections.  
 \\*

%--------------------------------------
\section*{Acknowledgments}

The author (Mahesh Choudhary) is thankful for financial support in the form of Senior Research Fellowships from the Council of Scientific and Industrial Research (CSIR), Government of India, (File No 09/013(882)/2019-EMR-1). The SERB, DST, Government of India [Grant No. CRG/2019/000360], IUAC-UGC, Government of India (Sanction No. IUAC/XIII.7 /UFR- 71353) and Institutions of Eminence (IoE) BHU [Grant No. 6031] are also gratefully acknowledged by one of the author (A. Kumar).\\
We acknowledge the kind support provided by Prof. Chandana Bhattacharya, Head, Experimental Nuclear Physics Division, VECC, Kolkata and Prof. A. K. Tyagi, Director, Chemistry Group, BARC, Mumbai towards the successful execution of the experiment. We would also like to express our gratitude to VECC Cyclotron (K-130) staff for providing us with high-quality beams throughout the experiment.


\begin{thebibliography}{99}

\bibitem{03} B. Mukhopadhyay and K Mukhopadhyay,   J. Nucl. Med. Radiat. Ther. \textbf{2}(2) (2011) 1000115. 
\bibitem{04} A. A. Alharbi \textit{et al.}, In Radioisotopes-Applications in Bio-Medical Science. IntechOpen, (2011).
\bibitem{another} R. Avagyan \textit{et al.}, Universal Journal of Applied Science \textbf{2}(7) (2014) 221.
\bibitem{13}  N. Otuka \textit{et al.}, Nucl. Data Sheets \textbf{120} (2014) 272.
\bibitem{14} IAEA-EXFOR Experimental nuclear reaction database, https://www-nds.iaea.org/exfor (Data retrieved on January 2023).
\bibitem{Mahesh} M. Choudhary \textit{et al.}, The European Physical Journal A \textbf{58} (2022) 95.

\bibitem{10} A. Gandhi \textit{et al.}, European Physical Journal A \textbf{57} (2021) 1.


\bibitem{16} A. J. Koning, S. Hilaire and M. C. Duijvestijn, TALYS-1.0, Proceedings of the International Conference on Nuclear Data for Science and Technology, April 22-27, 2007, Nice, France, editors O. Bersillon, F. Gunsing, E. Bauge, R. Jacqmin, and S. Leray, EDP Sciences, (2008) 211.



\bibitem{cy1} A. Goswami \textit{et al.}, Pramana \textbf{93} (2019) 1-13.



\bibitem{17} M. S. Uddin, K. S. Kim, M. Nadeem, S. Sudar,  and G. N.  Kim,  The European Physical Journal A \textbf{53}(5) (2017) 1-10.
\bibitem{18} S. Takacs, M. P. Takacs,  A. Hermanne,  F. Tarkanyi,  and R. A. Rebeles, Nuclear Instruments and Methods in Physics Research Section B: Beam Interactions with Materials and Atoms \textbf{297} (2013) 44-57.
\bibitem{19} S. Takacs, M. P. Takacs, A. Hermanne, F. Tarkanyi, and R. A. Rebeles, Nuclear Instruments and Methods in Physics Research Section B: Beam Interactions with Materials and Atoms \textbf{278} (2012) 93-105.
\bibitem{20} T.  Siiskonen, J. Huikari, T. Haavisto, J. Bergman, S. J. Heselius, J. O. Lill,  T. Lonnroth and K. Perajarvi,  Applied Radiation and Isotopes \textbf{67}(11) (2009) 2037-2039.


\bibitem{21} J. F. Ziegler,  J. P. Biersack and M.  D. Ziegler 2018 SRIM-The Stopping and Range of Ions in Matter (SRIM Co., 2008). URL: http://www. SRIM. org. 
\bibitem{22} Peter Sigmund, and Schinner Andreas, Nuclear Instruments and Methods in Physics Research Section B: Beam Interactions with Materials and Atoms \textbf{410} (2017) 78-87.

\bibitem{23} L. R. M Punte, B. Lalremruata, N. Otuka, S. V. Suryanarayana, Y. Iwamoto, R. Pachuau, B. Satheesh, H. H. Thanga,  L. S. Danu, V. V. Desai, L. R. Hlondo, S. Kailas, S. Ganesan, B. K. Nayak, A. Saxena, Phys. Rev. C \textbf{95} (2017) 024619.

\bibitem{24} H. Rameback \textit{et at.}, Journal of Radioanalytical and Nuclear Chemistry \textbf{304(1)} (2015) 467-471.
\bibitem{25} T. Vidmar, G. Kanisch, and G. Vidmar, App. Radiat. Isot. \textbf{908} (2011) 69.
\bibitem{26} R. Pachuau \textit{et al.}, Nucl. Phys. A \textbf{992} (2019) 121613.
\bibitem{28} D. L. Smith, and N. Otuka, Nuclear Data Sheets \textbf{113}(12) (2012) 3006-3053.
\bibitem{29} B. Lawriniang \textit{et al.}, Journal of Radioanalytical and Nuclear Chemistry \textbf{319}(3) (2019) 695-701.
\bibitem{30} N. Otuka \textit{et al.}, Radiation Physics and Chemistry \textbf{140} (2017) 502-510.
\bibitem{15} A. J. Koning and D. Rochman, Nucl. Data Sheets \textbf{113} (2012) 2841.
\bibitem{31} Mert Sekerci, Radiochimica Acta \textbf{108}(6) (2020) 459-467.

\bibitem{35} A. Gilbert and A. G. W. Cameron, Canadian Journal of Physics \textbf{43} (1965) 1446.
\bibitem {36} W. Dilg, W. Schantl, H. Vonach, and M. Uhl, Nucl. Phys. A \textbf{217} (1973) 269.

\bibitem {37} A. V. Ignatyuk, J. L. Weil, S. Raman, and S. Kahane, Phys. Rev. C \textbf{47} (1993) 1504.
\bibitem{38} S. Goriely, F. Tondeur, and J. M. Pearson, A Hartree–Fock nuclear mass table, Atomic Data and Nuclear Data Tables \textbf{77}(2) (2001) 311-381.
\bibitem{39} S. Goriely, S. Hilaire and A. J. Koning, Phys. Rev. C \textbf{78} (2008) 064307.
\bibitem{40} S. Hilaire, M. Girod, S. Goriely, and A. J. Koning,  Phys. Rev. C \textbf{86} (2012) 064317.
\bibitem{41}  A. Karpeles,  Radiochimica Acta  \textbf{12} (1969) 115-117.
\bibitem{42} Y Nagame, \textit{et al.}, The International Journal of Applied Radiation and Isotopes \textbf{29} (1978) 615-619.
\bibitem{43} M. Aikawa, M. Saito, S. Ebata, Y. Komori and H. Haba, Nuclear Instruments and Methods in Physics Research Section B: Beam Interactions with Materials and Atoms \textbf{427} (2018) 91-94.
\end{thebibliography}
\end{document}